\documentclass[page-classic]{epl2}
\usepackage{graphicx}
\usepackage{wasysym}
\usepackage{amssymb}

\begin{document}
\title{Corrections to scaling in the dynamic approach to the phase transition with quenched disorder}
\shorttitle{Corrections to scaling with quenched disorder}

\author{L. WANG\inst{1}, N. J. ZHOU\inst{2} \and B. ZHENG\inst{1}\footnote{corresponding author; email: zheng@zimp.zju.edu.cn}}
\institute{\inst{1}
  Department of Physics, Zhejiang University, Hangzhou 310027, P. R. China\\ \inst{2} Department of
  Physics, Hangzhou Normal University, Hangzhou 310036, P. R. China}
\shortauthor{L. WANG, N. J. ZHOU \and B. ZHENG}

\abstract{With dynamic Monte Carlo simulations, we investigate the continuous phase transition in the
  three-dimensional three-state random-bond Potts model.
  We propose a useful technique to deal with the strong corrections to the dynamic scaling form.
  The critical point, static exponents $\beta$ and
  $\nu$, and dynamic exponent $z$ are accurately determined. Particularly,
  the results support that the exponent $\nu$ satisfies the lower bound $\nu \geqslant 2/d$.}

  \maketitle

\section{Introduction}
The effects of quenched disorder on phase transitions are
of great interest in physics for decades. Recent efforts on the theoretical and numerical
studies can be found for example in \revision {Refs.~\cite{dot95,car97,jac98,cha05,ken08,gre09,fyt10,gre10,bal10,mur11,the11,fer12,bel12,mal12,mur12,xio12}.}
It has been rigorously proven that an infinitesimal amount of quenched disorder coupled to the energy density will turn a
first-order transition to a continuous one when the spatial dimension $d\leqslant 2$ \revision {\cite{aiz89,hui89}}. This
phenomenon has been observed, for example, in the two-dimensional ($2$D) eight-state Potts
model, Blume-Capel model and three-Color
Ashkin-Teller model\cite{che92,mal09,bel12}. For the case of $d=3$, the phase transition may persist first order if the strength of
quenched disorder is weak, and then soften to continuous if the strength is strong enough \cite{cha01,yin05,mur11,mal12}.

The effects of quenched disorder, imposed to a pure system with a continuous phase transition, can be judged by
the specific heat exponent $\alpha_p$ or the correlation length exponent $\nu_p$ of the pure system according to
the Harris criterion\cite{har71}. \revision {If $\alpha_p<0$ or $\nu_p>2/d$, the disorder is irrelevant to the
critical behavior, and if $\alpha_p >0$ or $\nu_p < 2/d$,} the disorder is relevant, and leads to a new universality class governed by
the `disorder' fixed point\cite{fyt10,the11}.
\revision {For the marginal case $\alpha_p=0$, the criterion can not give a conclusion whether the disorder is relevant or not, and numerical results support that the model obeys the strong universality hypothesis with logarithmic correlations\cite{ken08,fyt08}.
On the other hand, for a disordered system with a continuous phase transition, which is governed by the `disorder' fixed point,}
the specific heat exponent $\alpha$ does not decide whether the disorder is relevant.
It is only proven that a lower bound $\nu \geqslant 2/d$ should be satisfied\cite{aha78,kin81,cha86}.

To clarify the effects of quenched disorder on phase transitions, extensive numerical studies have been performed, for example,
for the three-dimensional (3D) $q$-state site-diluted Potts
model, bond-diluted Potts model and random-bond Potts
model\cite{cha05,yin05,mur12,xio12,fer12}. Recently, a numerical study based on the finite-time scaling \revision {and
dynamic Monte Carlo renormalization-group methods} gives $\nu=0.554(9)$ for the 3D three-state
random-bond Potts model\cite{xio12}, which violates $\nu \geqslant 2/d$. Such a result is rationalized as
the intrinsic $\nu$\cite{paz97}, which might
differ from the one measured from the finite-size scaling. This
controversy prompts us to clarify the problem with alternative methods.

In the past years, much progress has been achieved in understanding the dynamic relaxation processes.
For example, a pioneer work by Janssen et al yields a universal dynamic
scaling form, which is valid even at the macroscopic short-time regime\cite{jan89}.
Base on the dynamic scaling form, a series of methods and techniques have been developed to measure the critical point, and the critical exponents
including the static and dynamic ones\cite{zhe98,luo98,zhe99,alb11}.
Since the measurements are carried out in the short-time regime, the methods do not suffer from critical slowing down.
A variety of important problems can be tackled with the short-time dynamic approach \cite{zhe98,luo98,zhe99,alb11}, including very recent ones such as the interface growth, domain-wall motion and structural glass dynamics\cite{zho09,zho10,bus12,qin12,don12a,gro13,tor13,zho13}.

According to Ref. \cite{paz97}, for a statistical system with quenched disorder,
the effective critical point for a finite lattice $L^d$ should be dependent of the disorder
realization. For example, in the finite-size scaling form of a physical quantity in the equilibrium state,
\revision {$\langle Q \rangle = L^{-y}Q(L^{1/\nu}\ \tau)$,}
the reduced temperature $\tau$ should be replaced by $\tau +\delta \tau$, and $\delta \tau$
is assumed to depend on the disorder realization in the form $\delta \tau \propto xL^{-d/2}$,
with $x$ being a random variable associated with the disorder realization.
Under such an assumption, even if $\tau=0$, $L^{1/\nu}\delta \tau$ may drive the system away from the critical point,
in case $\nu < 2/d$. This is supposed to be the origin why the exponent $\nu$ measured from the finite-size scaling form
may be different from the intrinsic one.

In fact, the short-time dynamic approach is a proper method to tackle the problem.
In the short-time regime of the dynamic evolution, the finite-size effect is negligibly small,
and the physical quantity is scaled as \revision {$\langle Q \rangle = \xi(t)^{-y}Q(\xi(t)^{1/\nu}\ \tau)$.}
Here $\xi(t)$ is the non-equilibrium spatial correlation length, typically characterized by the dynamic exponent $z$ in the form $\xi(t) \propto t^{1/z}$.
In the short-time regime, it is assumed $\xi(t)\ll L$.
Even if there may be a shift of the critical point,
$\xi(t)^{1/\nu}\delta \tau$ does not drive the dynamic relaxation away from the critical point.
\revision {However, the disorder may lead to corrections to scaling.}
In the short-time dynamic approach, the critical point is usually determined by searching for the
best power-law behavior of a physical observable, e.g., the magnetization $\langle M \rangle \propto t^{-\beta/\nu z}$. Therefore, a strong correction to scaling conceptually and
significantly reduces the reliability and accuracy in the determination of the critical point,
and this naturally affects the measurements of the critical exponents.
In fact, a strong correction to scaling is often observed\cite{yin05,lei07,ken08,had08,for13}.

In this article we propose a useful technique to deal with the corrections to
scaling, and determine the critical point accurately. Our main idea is to separate the correction to scaling of the
spatial correlation length $\xi(t)$ from that of the scaling function of the physical observable.
Usually, the former may be strong.
Since the power-law behavior $\langle Q \rangle \propto \xi(t)^{-y}$ is free of the correction to scaling of $\xi(t)$, we can use it for the determination of the critical point. With the accurate critical point at hand, we may measure
the static exponents $\beta$ and $\nu$, and the dynamic exponent $z$, and identify the
correction to scaling of $\xi(t)$.

\section{Model and scaling analysis}
\label{model}
The 3D $q$-state random-bond Potts model is defined by the Hamiltonian
\begin{equation}
  -\frac{1}{k_{B}T}H=\sum_{\langle ij\rangle} K_{ij} \delta_{\sigma_{i},\sigma_{j}}
  \mbox{, }K_{ij}>0\mbox{.}
\end{equation}
Here, $\delta_{\sigma_{i},\sigma_{j}}$ is the Kronecker delta function, the spin $\sigma_i$ takes $q$
values from $1$ to $q$. $\langle ij\rangle$ denotes the nearest neighbor on a cubic lattice. The couplings $K_{ij}$ take two possible values,
$K_{1}=K$ and $K_{2}=rK$, with a bimodal distribution,
\begin{equation}
  P(K_{ij}) = p \delta(K_{ij} - K_{1}) + (1-p) \delta(K_{ij} - K_{2}).
\end{equation}
Here the ratio $r$ is the disordered amplitude, and $p$ is the probability for $K_{ij}=K_1$.
\revision {For $q=3$ at $r=1$, i.e., the pure three-state Potts model}, the phase transition is of first order.
For a large $r$, it is softened to the second order.
In this article, we perform the simulation for $q=3$, at $r=10$ and
$p=0.5$. The periodic boundary condition is employed for all the spatial directions.

The measured physical observables are the magnetization and its second momemnt,
\begin{equation}
  M^{(k)}(t)=\frac{q^{k}}{(q-1)^{k}L^{d k}} \left\langle \left[ \sum_{i} \left(
    \delta_{\sigma_{i}(t),1} - 1/q \right) \right]^{k} \right\rangle, \quad k=1,2
\end{equation}
where $d=3$ is the spatial dimension, $L$ denotes the lattice size, $\langle\cdots\rangle$ represents both the
statistical average and the average for disorder realizations, and $M^{(1)}(t)\equiv M(t)$ is just the magnetization.

After a microscopic time scale $t_{mic}$, a dynamic scaling form near the critical point is assumed\cite{jan89,zhe98,luo98,zhe99,alb11}
\begin{equation}
  M^{(k)}(t, \tau, L) = b^{-k\beta/\nu} M^{(k)}(b^{-1}\xi(t), b^{1/\nu}\tau, b^{-1}L),
  \label{scaling_1}
\end{equation}
where $b$ is an arbitrary scaling factor, $\tau$ is the reduced temperature, $\beta$ and $\nu$ are the static
critical exponents, and $\xi(t)$ is the non-equilibrium spatial correlation
length. Such a dynamic scaling form holds already from the macroscopic short-time regime\cite{jan89,zhe98,luo98,zhe99,alb11}.
For a large lattice, i.e., $\xi(t) \ll L$, the finite-size effect is negligibly small,
and the scaling form of the magnetization can be written as
\begin{equation}
\revision {  M(t,\tau) = \xi(t)^{-\beta/\nu } M(\xi(t)^{1/\nu} \tau ).}
  \label{scaling_Mxi}
\end{equation}
At the critical point, i.e., $\tau =0$, the magnetization obeys a power law $M \sim \xi(t)^{-\beta/\nu }$.
Away from the critical point, the scaling function $M(\xi(t)^{1/\nu} \tau)$ modifies the power-law behavior.
According to Eq.(\ref{scaling_Mxi}), the critical exponent $\nu$ can be identified from the derivative of $\ln M(t,\tau)$,
\begin{equation}
  \partial_{\tau}\ln M(t,\tau)|_{\tau=0} \sim \xi(t)^{1/\nu }.
  \label{plnM_o_ptau}
\end{equation}
In the scaling regime, the spatial correlation length $\xi(t)$ usually grows in a power-law form
\begin{equation}
\xi(t) \sim t^{1/z},
\label{tscaling}
\end{equation}
where $z$ is the dynamic critical exponent.
Inserting the above power-law growth of $\xi(t)$ into Eq.~(\ref{scaling_Mxi}), plotting the magnetization $M(t,\tau)$ as a function of the time $t$ and
searching for the best power-law behavior, one may determine the critical point and the critical exponent $\beta/\nu z$.
The critical exponent $1/\nu z$ is then extracted from Eq.~(\ref{plnM_o_ptau}).
This is the standard procedure in the dynamic approach to the second-order phase transition.

When there are strong correlations to scaling, however, the above standard procedure suffers.
For example, with a logarithmic correlation to scaling, it is in principle not possible to determine the critical point
based on Eq.~(\ref{tscaling}) in the short-time regime.
In fact, there are two kinds of corrections to scaling.
\revision {One is the correction of the power-law growth of the spatial correlation length $\xi(t)$ in Eq.~(\ref{tscaling}),
and another is that of the scaling function of the physical observable such as in Eq.~(\ref{scaling_Mxi}).}
Usually, the former can be strong, even in a logarithmic form.
To circumvent the difficulty induced by the correction to scaling of $\xi(t)$, our idea is to plot the magnetization $M(t,\tau)$
as a function of $\xi(t)$. Searching for the best power-law behavior, one may determine the critical point
and measure the critical exponents.

Now the problem is how to measure the spatial correlation length $\xi(t)$.
For this purpose, we introduce a Binder cumulant $U=M^{(2)}/M^{2}-1$.
\revision {From the dynamic scaling form in Eq.~(\ref{scaling_1}), one could deduce that the Binder cumulant scales as $U=U(\xi(t)/L,\xi(t)^{1/\nu} \tau )$.
Taking into account $\xi(t) \ll L$, one of the two summations over the lattice sites in $M^{(2)}$ is not extended to the whole lattice,
and it leads to $U \sim L^{-d}$ \cite{zhe98}. Thus,}
the scaling form of the Binder cumulant is written as
\begin{equation}
  U(t) = (\xi(t)/L)^{d}G(\xi(t)^{1/\nu} \tau ).
  \label{uxi}
\end{equation}
Inserting the above relation into Eqs.~(\ref{scaling_Mxi}) and (\ref{plnM_o_ptau}), we obtain
\begin{equation}
  M(t,\tau) = U(t)^{-\beta/\nu d} F(U(t)^{1/\nu d} \tau)
  \label{eq-mu}
\end{equation}
and
\begin{equation}
  \partial_{\tau}\ln M(t,\tau)|_{\tau=0} \sim U(t)^{1/\nu d}.
  \label{eq-dmu}
\end{equation}
\revision {Here the argument $U(t)$ should actually be $L^d U(t)$.
The factor $L^d$ is ignored, or absorbed into the scaling function, just for convenience in writing.}
Finally, the dynamic exponent $z$ and the correction to scaling of $\xi(t)$ can be extracted from the Binder cumulant $U(t)$.

\section{Monte Carlo simulations}
With the Metropolis algorithm, we preform dynamic Monte Carlo simulations starting from an ordered initial state, to study
the phase transition in the 3D three-state
random-bond Potts model. The disorder amplitude is taken to be $r=10$, and main results are presented with the lattice size $L=150$. Additional
simulations with $L=100$ and 200 confirm that the finite-size effect is already negligibly small. The number of
samples for averaging over the disorder realizations is up to 24000. The updating time is $10000$ Monte Carlo
steps. A Monte Carlo step is defined as a sweep over all the lattice sites. The total samples are divided into three
subgroups to estimate the statistical errors. Further, the fluctuations of the measurements
in different time intervals will be also taken into account if
comparable to the statistical ones. In particular, corrections to scaling are carefully analyzed.
\revision {According to standard renormalization-group calculations, we typically consider the corrections in a power-law form, i.e., $(1+c/t^b)$.
As the correction exponent $b$ goes to zero, it becomes stronger and stronger, and finally approaches the logarithmic one, i.e., $(1+c \ln t)$.
}

\begin{figure}[h]
  \begin{center}
    \scalebox{0.35}{\includegraphics{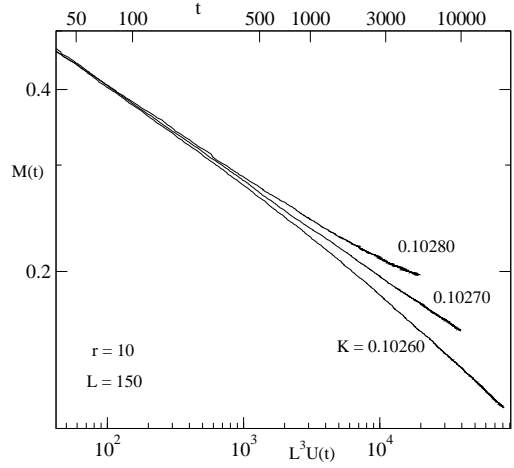}}
  \end{center}
  \caption{The magnetization $M(t)$ vs. the Binder cumulant $L^{3}U(t)$ at $r=10$
    is plotted on a log-log scale. The critical point is determined to be $K_{c}=0.10270(1)$.
    The upper $t$-scale corresponds to that of $L^{3}U(t)$ at K=0.10270.}
  \label{ptp3}
\end{figure}

In Fig.~\ref{ptp3}, we plot the magnetization $M(t)$ as a function
of the Binder cumulant $U(t)$ for different inverse temperatures $K$. \revision {For
comparison of different lattice sizes in next steps, the rescaled Binder cumulant $L^{3}U(t)$ is used as the
argument. For different $K$'s, the time evolution of the Binder cumulant $U(t)$ is different. The time scale in the upper $x$-axis is for $K=0.10270$.
To locate the critical point, we linearly interpolate the magnetization $M(t)$ between $K=0.10260$, $0.10270$ and $0.10280$
with $\Delta K=0.00001$. Extra direct simulations also confirm that such a linear interpolation is already very accurate.
Skipping the data in a microscopic time scale, typically $t_{mic} > 30$, fitting the curve of $M(t)$ at each $K$ to a power law,
and searching for the best power-law behavior of $M(t)$,
we determine the critical point $K_{c}=0.10270(1)$ \cite{zhe98,luo98,zhe99,alb11}. The error is estimated by dividing the total samples into three subgroups.}
Within statistical errors, our measurement of $K_c$ is compatible with
$K_{c}=0.10265(5)$ in Refs.~\cite{yin05} and $K_{c}=0.1026(4)$ in Ref.~\cite{xio12}, but with a higher accuracy.
Such a higher accuracy is important for the measurements of the critical exponents.

\begin{figure}[h]
  \begin{center}
    \scalebox{0.35}{\includegraphics{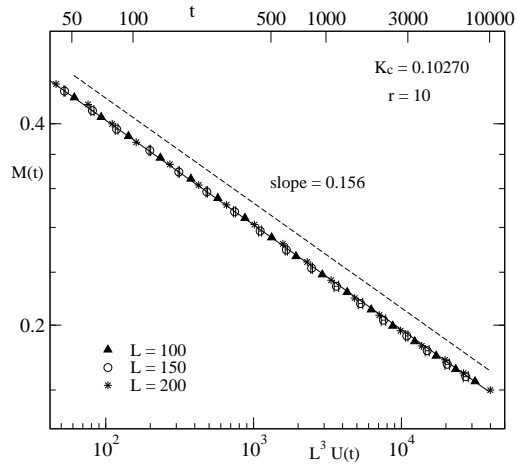}}
  \end{center}
  \caption{The magnetization $M(t)$ vs. the Binder cumulant $L^{3}U(t)$ at $K_{c}$ for
    different lattices is plotted on a log-log scale. The dashed line represents a power-law fit and the solid line is with a power-law correction.
    The error bars of $L^{3}U(t)$ for $L=150$ are shown, but they are smaller than the symbols.}
  \label{ptppow}
\end{figure}

In Fig.~\ref{ptppow}, the magnetization $M(t)$ is plotted as the function of
$L^{3}U(t)$ at the critical point $K_{c}$. Based on Eq.~(\ref{eq-mu}), the exponent $\beta/\nu=0.468(1)$ is then
extracted from the slope of the curve.
\revision {Following Ref.~\cite{cha01}, one may perform the $\chi^2$ test for the power-law fit of the curve.
The value of $\chi^2/DoF$ is about $1$ when skipping the data of $t \leqslant 200$.
The curve at early times yields the value of $\chi^2/DoF$ up to $3$, possibly due to the very small fluctuations of the data points.}
One may consider the correction to scaling, for example, $M \sim U^{\beta/\nu d}(1+c / U^{b})$.
Such a fit to the numerical data is indeed better than a simple power law, and yields $b=2.1$ and $c=-5.2$.
But the exponent $\beta/\nu$ remains almost unchanged.
In Fig.~\ref{ptppow}, we also show that the finite-size effect is already negligibly small
for $L=100$, $150$ and $200$.

\begin{figure}[h]
  \begin{center}
    \scalebox{0.35}{\includegraphics{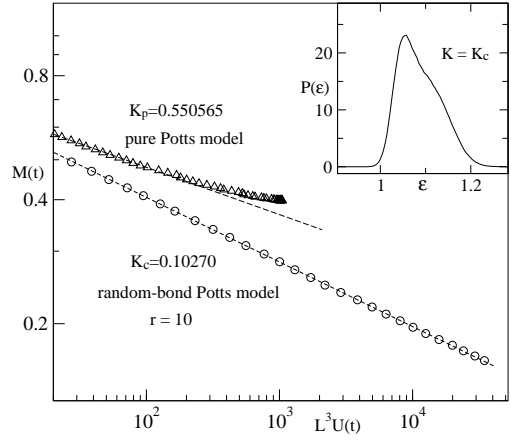}}
  \end{center}
  \caption{\revision {The magnetization $M(t)$ of the three-state random-bond Potts model at $K_c$ is compared with that of the
    pure three-state Potts model at the first-order transition point $K_p$ on a log-log scale.
    The transition point $K_{p}$ is taken from Ref.~\cite{jan97}. Dashed lines
    show power-law fits. In the inset, a single peak behavior of the probability distribution
    of the energy density is displayed for the three-state random-bond Potts model at $K_c$. }
    }
  \label{compare}
\end{figure}

\revision {The phase transition of the pure three-state Potts model (i.e., $r=1$) is of the first order.
With the strong random-bond disorder such as $r=10$, the phase transition is softened to the second order.
The power-law behavior of the magnetization for $r=10$ at the critical point $K_c$ already indicates such a softening phenomenon \cite{zhe98,yin05}.
As shown in Fig.~\ref{compare}, the dynamic behavior of the magnetization of the pure three-state Potts model
obviously deviates from a power law. In addition, we have performed simulations in equilibrium,
and measure the probability distribution of the energy density. In the inset of Fig.~\ref{compare},
a single peak structure for the random-bond Potts model at $r=10$ is clearly observed, which is a standard signal
of a second-order phase transition \cite{cha01,fyt08}.
}

\begin{figure}[h]
  \begin{center}
    \scalebox{0.35}{\includegraphics{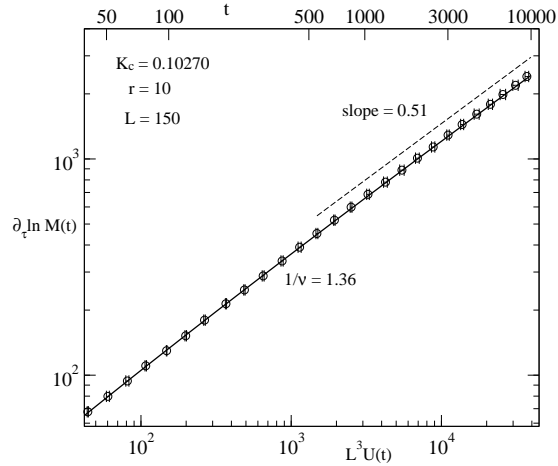}}
  \end{center}
  \caption{$\partial_{\tau}\ln M(t)$ vs. the Binder cumulant $L^{3}U(t)$ at $K_{c}$
    is plotted on a log-log scale. The dashed line represents a power-law fit and the solid line
    is with the logarithmic correction.}
  \label{lnM_ln}
\end{figure}

To measure the exponent $\nu$ with Eq.~(\ref{eq-dmu}), we calculate the logarithmic derivative of $M(t)$ in
the neighborhood of $K_{c}$.
As shown in Fig.~\ref{lnM_ln}, the slope of the curve continuously changes with time.
A direct measurement in the time regime $[500,10000]$ gives $1/\nu d=0.51$, i.e., $1/\nu=1.53$. But a correction to scaling does exist.
A power-law correction of the form $\partial_{\tau}\ln M(t, \tau) \sim
U^{1/\nu d}(1+ c/U^{b})$ yields $1/\nu=1.38(3)$ and $b=0.02$.
A small $b$ usually implies the existence of a logarithmic correction. We then fit the numerical data to
the form $ \partial_{\tau}\ln M(t, \tau) \sim U^{1/\nu d}(1+c\ln U)$, and obtain $1/\nu=1.36(1)$.
Both results support that $1/\nu \leqslant d/2$ or equivalently $\nu \geqslant 2/d$ should be satisfied.
\revision {We have also performed the $\chi^2$ test for both the power-law and logarithmic corrections, 
and the values of $\chi^2/DoF$ in the time interval $[50,10000]$ are about $0.4$.}

\begin{figure}[h]
  \begin{center}
    \scalebox{0.35}{\includegraphics{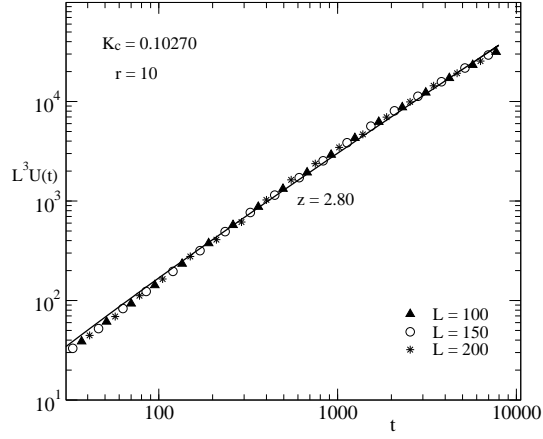}}
  \end{center}
  \caption{The Binder cumulant $U(t)$ at $K_c$ is plotted on a log-log scale. The solid line
    is a fit with the logarithmic correction.}
  \label{Binder_ln}
\end{figure}

According to Eq.~(\ref{uxi}), at $Kc$, the Binder cumulant $U(t) \sim (\xi(t)/L)^d$.
In Fig.~\ref{Binder_ln}, $U(t)$ is plotted at $K_{c}$ to explore the growth law
of the spatial correlation length $\xi(t)$. Obviously, a strong correction to scaling exists, and its behavior is somewhat complicated.
Basically, however, a logarithmic correction to Eq.~(\ref{tscaling}), i.e.,  $U(t) \sim t^{d/z}(1+c\ln t)$, fits to the data.
The exponent $d/z$ is estimated to be $1.07(1)$, and thus $z=2.80(3)$. \revision {To support our results, we have computed the spatial correlation
function $C(t,r)$, and directly extracted the spatial correlation length $\xi(t)$ from the exponential decay $C(t,r) \sim \exp (-r/\xi(t))$
in large-$r$ regimes.
The resulting $\xi(t)$ confirms the relation $U(t) \sim (\xi(t)/L)^d$.}
In the literatures, a similar logarithmic corrections to $U(t)$ is also detected in the 2D site-diluted Ising model
\cite{ken08,had08}. \revision {Further, for the dynamic relaxation at low temperatures, it has been found that the growth law $\xi(t)$
for disordered systems may cross over from the power law in the transient regime to the logarithmic one in the large-$t$ regime \cite{par10}.}

\begin{figure}[h]
  \begin{center}
    \scalebox{0.35}{\includegraphics{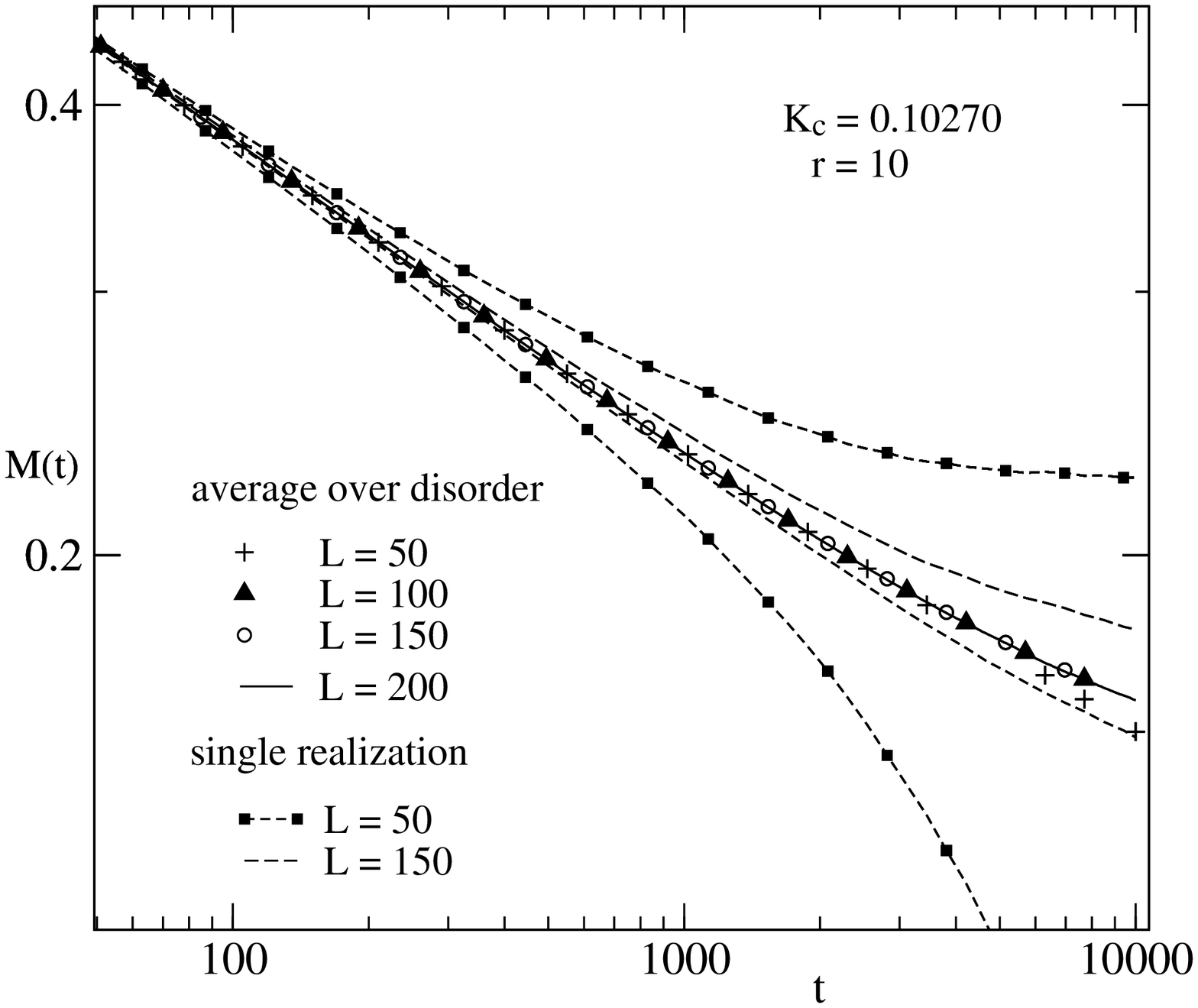}}
  \end{center}
  \caption{\revision {The dynamic evolution of the magnetization $M(t)$ averaged over different realizations of the random-bond disorder at $K_c$
  is plotted on a log-log scale, in comparison with that of a single realization. }
  }
  \label{realization}
\end{figure}

\revision {Due to the strong logarithmic correction to scaling of $\xi(t)$, the critical point $K_c=0.10270$ determined
from the power-law behavior $M(t) \sim \xi(t)^{-\beta/\nu}$ differs from the naive one obtained from $M(t) \sim t^{-\beta/\nu z}$. In fact, the latter is around $K=0.10265$ \cite{yin05}.
In Fig.~\ref{realization}, the magnetization $M(t)$ at $K_c=0.10270$ is plotted as a function of the time $t$ on a log-log scale.
A clear deviation from the power-law behavior is observed, which is induced by the logarithmic correction of $\xi(t)$.
In our computations, an average over different realizations of the random-bond disorder has been performed,
and no finite-size effect is detected for $L \geqslant 100$. In other words, such a logarithmic correction of $\xi(t)$ remains even for an infinite lattice.
In Fig.~\ref{realization}, it is also shown
that the magnetization $M(t)$ with a single realization of the random-bond disorder fluctuates around the curve
averaged over different realizations.
Such a fluctuation should be the order of $\xi(t)^{1/\nu}\delta \tau \sim \xi(t)^{1/\nu} xL^{-d/2}$
as described in the section of Introduction. In the short-time regime, this fluctuation is negligibly small
for a very large lattice, or equivalently, can be removed by the average over different realizations of the disorder.
In the equilibrium state, however, $\xi(t)$ will grow to the order of $L$. In case $\nu < 2/d$,
it may drive the system away from the critical point.
}

In Table \ref{table1}, we list the critical point and all the critical exponents we measure for the 3D three-state
random-bond Potts model. Compared with previous works, the critical point is
more accurate. Due to the slightly larger $K_c$, the static exponents $\beta/\nu$ and $1/\nu$ become smaller and the dynamic
exponent $z$ is also a little larger.
The value $1/\nu=1.36(1)$ strongly supports that $1/\nu \leqslant d/2$, in contrast to $1/\nu=1.80(4)$
obtained with the finite-time scaling \revision {and
dynamic Monte Carlo renormalization-group methods} in Ref\cite{xio12}.
In Ref\cite{yin05}, the logarithmic correction to the power-law growth of the spatial correlation length
$\xi(t)$ is ignored in the determination of the critical point, the numerical data are limited, and the statistical errors and corrections to scaling
in the measurements of the critical exponents are much larger.
\revision {In Table \ref{table2}, the critical exponents are compared for different statistical models with quenched disorder.
All the values of the exponent $\nu$ satisfy the lower bound $\nu \geqslant 2/d$, although some are marginal.}

\section{Conclusion}
With dynamic Monte Carlo simulations, we have investigated the continuous phase transition in the 3D three-state random-bond Potts model.
Due to the strong corrections to scaling induced by the quenched disorder, it is difficult to determine the critical point and critical
exponents accurately. We propose to separate the correction to scaling of the spatial correlation length
$\xi(t)$ from that of the scaling function $F(y)$ of the physical observable.
Based on the dynamic behavior of the magnetization $M(t)$ as a function of $\xi(t)$,
an accurate critical point $K_{c}=0.10270(1)$ is determined.
The static exponents $\beta$ and $\nu$, and the dynamic exponent $z$ are then extracted, and summarized in Table~\ref{table1}.
The results support that \revision {$\nu \geqslant 2/d$} should be satisfied.
The technique adopted in this article should be generally appropriate for dealing with the corrections to scaling
in continuous phase transitions.

\acknowledgements This work was supported in part by NNSF of China
under Grant Nos. 11375149 and 11205043, and Zhejiang Provincial
Natural Science Foundation of China under Grant No. LQ12A05002.


\begin{table}[h]
  \caption{The critical point and critical exponents for the 3D three-state random-bond Potts model with
    $r=10$, obtained from $M(t)$, $\partial_{t}\ln M(t,\tau)$ and $U(t)$. For comparison, the results of previous studies are listed as well.}
  \begin{tabular}{c|ccc}
    \hline
    & This work &Ref\cite{yin05}&Ref\cite{xio12}\\
    \hline
    $K_{c}$&$0.10270(1)$&$0.10265(5)$&$0.1026(4)$\\
    $\beta/\nu$&$0.468(3)$&$0.548(13)$&$0.54(6)$\\
    $1/\nu$&$1.36(1)$&$1.42(7)$&$1.80(4)$ \\
    $z$&$2.80(3)$&$2.48(5)$ &$2.51(4)$\\
    \hline
  \end{tabular}
  \label{table1}
\end{table}

\begin{table}[h]
  \caption{The critical exponents for the 3D three-state random-bond Potts (3-s RBP) model obtained in this article, in comparison with those of the three-state
    site-diluted Potts (3-s SDP) model, four-state bond-diluted Potts (4-s BDP) model,
    large-$q$-state random-bond Potts (LRBP) model, site-diluted Ising (SDI) model and random-bond Ising
    (RBI) model.}

  \begin{tabular}{c|ccc}
    \hline &$\beta$ &$\nu$ &$\beta/\nu$\\
    \hline
    3-s RBP ($p=0.5$) &$0.344(3)$ &$0.735(7)$& $0.468(1)$ \\
    3-s SDP ($p=0.65$) \cite{mur12} & $0.376(6)$ & $0.688(8)$ & $0.546(3)$ \\
    4-s BDP ($p=0.56$) \cite{cha05} & $0.547(39)$ &$0.747(19)$ & $0.732(24)$ \\
    LRBP ($p=0.5$) \cite{mer06} &  $0.45(2)$ & $0.73(1)$ & $0.60(2)$\\
    SDI ($p=0.8$) \cite{pru10} & $0.348(11)$ & $0.685(21)$ & $0.508(17)$ \\
    RBI ($p=0.5$) \cite{fyt10} & $0.357(19)$ & $0.688(15)$ & $0.519(9)$ \\
    \hline
  \end{tabular}
  \label{table2}
\end{table}

\newpage

\end{document}